\documentclass[journal, 10pt]{IEEEtran}
\usepackage{amsmath,amsfonts}
\usepackage{algorithmic}
\usepackage{algorithm}
\usepackage{array}
\usepackage{epsfig}
\usepackage{caption}
\usepackage[tight,footnotesize]{subfigure}
\usepackage{textcomp}
\usepackage{stfloats}
\usepackage{url} % 如果已经加载过则不用重复
\usepackage{hyperref}
\hypersetup{
    breaklinks=true,       % 允许链接中断行
    colorlinks=False,       % 可选：彩色链接
    urlcolor=False,         % 可选：链接颜色
}
\usepackage{verbatim}
\usepackage{graphicx}
\usepackage{cite}
\usepackage{xcolor}
\usepackage{balance}
\usepackage{booktabs}
\usepackage{placeins} 
\usepackage{multirow}

\begin{document}

\title{Environment Reconstruction in Terahertz Monostatic Sensing: Joint Millimeter-level Geometry Mapping and Material Identification}

\author{Zitong Fang, 
Yejian Lyu, Ziming Yu,
and Chong Han, \textit{Senior Member, IEEE}

\thanks{
This article was presented in part at the EuCAP 2025~\cite{fang2024centimeter}.

Zitong Fang (co-first author) and Yejian Lyu (co-first author) are with the Terahertz Wireless Communications (TWC) Laboratory, Shanghai Jiao Tong University, Shanghai, China (e-mail:
\{zitong.fang, yejian.lyu\}@sjtu.edu.cn).

Ziming Yu is with Huawei Technologies Co., Ltd, China (e-mail: yuziming@huawei.com).

Chong Han is with the Terahertz Wireless Communications (TWC) Laboratory and also the Cooperative Medianet Innovation Center (CMIC), School of Information Science and Electronic Engineering, Shanghai Jiao Tong University, China (Email:
chong.han@sjtu.edu.cn).
}% <-this % stops a space
}



\maketitle

\begin{abstract}
Terahertz (THz) integrated sensing and communication (ISAC) offers high-speed communication alongside precise environmental sensing. This paper presents a computationally efficient framework for THz-based environment reconstruction by integrating connected component analysis (CCA)-assisted multipath component (MPC) estimation with a sliding-window refinement strategy. To start with, a monostatic sensing experiment is conducted in an indoor scenario using a vector network analyzer (VNA)-based sounder operating from 290 to 310~GHz. On one hand, as for geometry mapping, a CCA-based region search is employed to accelerate parameter extraction, significantly reducing the search space for space-alternating generalized expectation-maximization (SAGE)-based estimation and achieving an 8.4 times acceleration, while preserving resolution. Further analysis of the connected component structure enables the identification of indoor features such as flat walls and corners. A sliding-window refinement applied to the identified regions improves geometric mapping, achieving the mean distance error of 4.9~mm, which is one order of magnitude better than the literature. On the other hand, the deterministic and stochastic components of the monostatic channel are classified through reflection loss analysis. Then, material identification is performed by looking up the reflection loss in a THz time-domain spectroscopy (THz-TDS) database, which comprises over 200 materials across a 0-6~THz range. Experimental results validate millimeter-level accuracy in geometry mapping and reliable material classification, enhancing the environmental awareness capabilities of THz ISAC systems.
\end{abstract}



\maketitle

\section{INTRODUCTION}
\IEEEPARstart{T}{he} rapid evolution of sixth-generation (6G) and beyond wireless networks is accelerating the development of integrated sensing and communication (ISAC) systems, which aim to unify communication and environmental perception functionalities~\cite{zhang2023integrated,11048955,isac_low_freq_model}. To support such advanced ISAC capabilities, the exploration of the terahertz (THz) band (0.1–10~THz) has gained significant attention due to its potential for ultra-high-speed data transmission, fine-grained sensing, and ubiquitous connectivity~\cite{9585685,akyildiz2014terahertz,he2018design}. Beyond enabling high-capacity communication, THz technology also enhances advanced environmental sensing~\cite{dupleich2024characterization,wu2021thz}. The incorporation of THz communication and sensing facilitates high-resolution geometry mapping by leveraging the wide bandwidth in the THz band~\cite{chen2023scatterer,han2024thz}, which contributes to improved environmental reconstruction accuracy and supports emerging functions such as material identification in real-world scenarios.

%\IEEEPARstart{T}{he} rapid evolution of sixth-generation (6G) and beyond wireless networks is driving the exploration of the terahertz (THz) band (0.1-10~THz) to satisfy the increasing demand for ultra-high-speed data transmission, enhanced sensing capabilities, and ubiquitous connectivity~\cite{9585685,akyildiz2014terahertz,he2018design}. Beyond enabling high-capacity communication, the THz technology also enhances advanced environmental sensing, thereby advancing the development of integrated sensing and communication (ISAC) systems~\cite{dupleich2024characterization,wu2021thz}.
%The incorporation of THz communication and sensing facilitates high-resolution geometry mapping by exploiting the wide bandwidth in the THz band~\cite{chen2023scatterer,han2024thz}, which assist in the enhancement of environmental reconstruction precision and enable new function of material identification in real-world scenarios.
 
Accurate positioning and localization play a fundamental role in sensing and environmental reconstruction. Recent works have explored various localization techniques, leveraging millimeter wave (mmWave) and THz systems to achieve sub-meter precision. For instance, a map-free indoor localization method based on an ultrawideband large-scale array system was proposed in~\cite{ji2018map}, where a high-resolution maximum likelihood estimation (MLE) algorithm was applied to extract the delay and angle-of-arrival (AoA) of multipath components (MPCs), achieving a localization error below 0.3~m. Similarly, a virtual anchor node (VAN)-based localization framework was introduced in~\cite{yassin2018mosaic}, which jointly estimates the receiver position and environmental obstacles. In~\cite{rastorgueva2024millimeter}, a snapshot simultaneous localization and mapping (SLAM) method was presented for mmWave localization at 60~GHz, using singular value decomposition (SVD)-based AoA/angle-of-departure (AoD) estimation, achieving a localization accuracy of 0.5~m.

Beyond localization, geometry reconstruction is a key component of THz-based sensing, enabling precise environmental mapping. 
Unlike microwave and mmWave systems, which suffer from limited spatial resolution, THz communication, with its expansive untapped spectrum, offers unprecedented data rates, opening up novel applications in high-resolution imaging and sensing~\cite{wei2022toward}. THz systems can exploit their short wavelengths and broad bandwidths to achieve centimeter-level or even millimeter-level accuracy~\cite{zhang2023channel}. A handful of state-of-the-art measurement-based ISAC research has been conducted. For instance,
the authors in~\cite{li2023300} conducted mapping experiments in three indoor scenarios between 306 and 321~GHz, where an average error of 0.1~m between the reconstructed geometry and the ground truth is observed. The authors in~\cite{lotti2023radio} proposed a radio SLAM algorithm for indoor scenarios to derive the map of laboratory/office and infer users' trajectories in the frequency range of 235-320~GHz. 
Recently, an environment reconstruction methodology leveraging multi-target reflector merging was proposed in~\cite{chang2025environment}, achieving a reconstruction error of 0.03~m in ray-tracing simulation and root-mean-square error (RMSE) of 1.28~m and 0.45~m in line-of-sight (LoS) and non-line-of-sight (NLoS) conditions, respectively. However, despite these advancements, current geometry mapping methods achieve decimeter-level accuracy that is limited to degraded precision in complex regions such as corners. 

In addition to geometry mapping, material identification is regarded the other essential part of environment reconstruction in ISAC systems, which is, however, commonly missing in the literature~\cite{jiang2024electromagnetic}. Uniquely in the THz band, different materials exhibit unique electromagnetic (EM) properties. Previous studies have utilized THz time-domain spectroscopy (THz-TDS) to investigate materials such as double-pane windows and painted plaster, measuring reflection properties from 100 to 500~GHz~\cite{jansen2008impact}. 
Moreover, angle- and frequency-dependent
measurements, coupled with Kirchhoff scattering theory, have
been used to model rough surface materials from 0.1 to 1~THz~\cite{jansen2011diffuse}. The absorption and reflection characteristics of typical building materials at various angles from 100 to 350~GHz are measured using THz-TDS,
as presented in~\cite{kleine2005characterization}. Although some efforts have been made to characterize material properties in the ISAC system using THz-TDS, practical measurements for material identification of real environment reconstruction are still lacking. 

To overcome challenges from limited accuracy in geometry and the lack of obtaining material information, we fill the research gap of THz environment reconstruction by achieving joint millimeter-level geometry mapping and reliable material identification. In our prior work~\cite{fang2024centimeter}, we proposed a two-dimensional (2D) space-alternating generalized expectation-maximization (SAGE) algorithm to jointly estimate the delay and angle of MPCs, along with a geometry-based method to mitigate the corner effect. However, the previous method suffers from high computational complexity due to the exhaustive parameter search and degraded accuracy in complex areas due to antenna pattern mismatch. In this work, we incorporate connected component analysis (CCA)-assisted connected component segmentation to reduce computational complexity, and employ a geometry-driven element-wise SAGE algorithm to mitigate antenna pattern mismatch. In addition, we propose a structure-aware sliding-window refinement strategy to further enhance mapping accuracy.
%First, monostatic sensing experiments are conducted in a laboratory environment using a vector network analyzer (VNA)-based channel sounder with a 20~GHz bandwidth spanning 290–310~GHz. Second, to efficiently estimate MPCs, a CV-assisted connected component analysis (CCA) is applied to segment dominant MPC regions, significantly reducing the search space for subsequent SAGE-based parameter estimation. The segmented connected component regions are further analyzed to infer key structural features of the indoor environment, such as flat walls and corners. Third, to enhance geometric mapping, a sliding-window refinement strategy is applied to the identified structures, effectively mitigating mapping errors. Furthermore, a THz-TDS material database comprising over 200 samples is utilized for reflection loss-based material identification. 
The key contributions of this paper can be summarized as follows:

\begin{itemize}
    \item \textbf{20~GHz bandwidth monostatic measurement in the THz band:} 
    Monostatic sensing experiments are conducted in a laboratory environment using a vector network analyzer (VNA)-based channel sounder spanning from 290 to 310~GHz over 10 monostatic locations, which corresponds to a distance resolution of 1.5~cm. 
    \item \textbf{Computationally efficient CCA-assisted MPC estimation:} 
    A CCA-based region search approach is applied to segment dominant MPC regions from the power-angle-delay profile (PADP). By reducing the search space of SAGE-based parameter estimation to these high-energy regions, the proposed framework achieves an 8.4 times acceleration in computation time compared to the previous element-wise SAGE algorithm in~\cite{mpc_estimation}, while maintaining high-resolution parameter estimation.

    \item \textbf{Structural inference and high-precision geometry mapping:} 
    The structure of connected component regions are further analyzed and identified to infer key indoor features such as flat walls and corners. Additionally, a sliding-window refinement technique is applied to improve mapping continuity and reduce geometric distortion in the reconstructed environment. These improvements lead to a millimeter-level reconstruction accuracy of 4.9~mm, at least one order of magnitude better than that reported in the literature.

    \item \textbf{Reliable material identification through reflection loss analysis:} 
    A material database is constructed based on THz-TDS, covering over 200 common materials in the 0–6~THz range. For each region extracted from channel measurements, reflection loss at 300~GHz is computed and used to distinguish deterministic and stochastic MPCs. By comparing the minimum reflection loss values with the database, the proposed method successfully identifies materials as cement for wall structure and metal for window frame structure, demonstrating practical material sensing capability.    
\end{itemize}

The remainder of this paper is structured as follows. Section II describes the monostatic THz measurement setup and experimental configuration. Section III presents the proposed CCA-assisted SAGE algorithm. Section IV discusses the methodology for geometric mapping and material identification. Section V provides an analysis of computational complexity, geometry mapping error, and material identification. Finally, Section VI concludes the paper.

\section{Measurement setup and scenario}
%In this section, the monostatic sensing experiments are described, including the sounding system and measurement scenario.

% \begin{figure}
% \centering
% \subfloat[THz measurement system and setup.]{\includegraphics[width=0.8\columnwidth]{figures/real.png}%
% \label{fig:scenario1}}
% \hfil
% \subfloat[Schematic of the scenario.]{\includegraphics[width=0.8\columnwidth]{figures/scenario.png}%
% \label{fig:scenario}}
% \caption{Measurement system and scenario.}
% \label{fig:description}
% \end{figure}
\begin{figure}
\centering
\subfigure[]{
\includegraphics[width=0.36\textwidth]{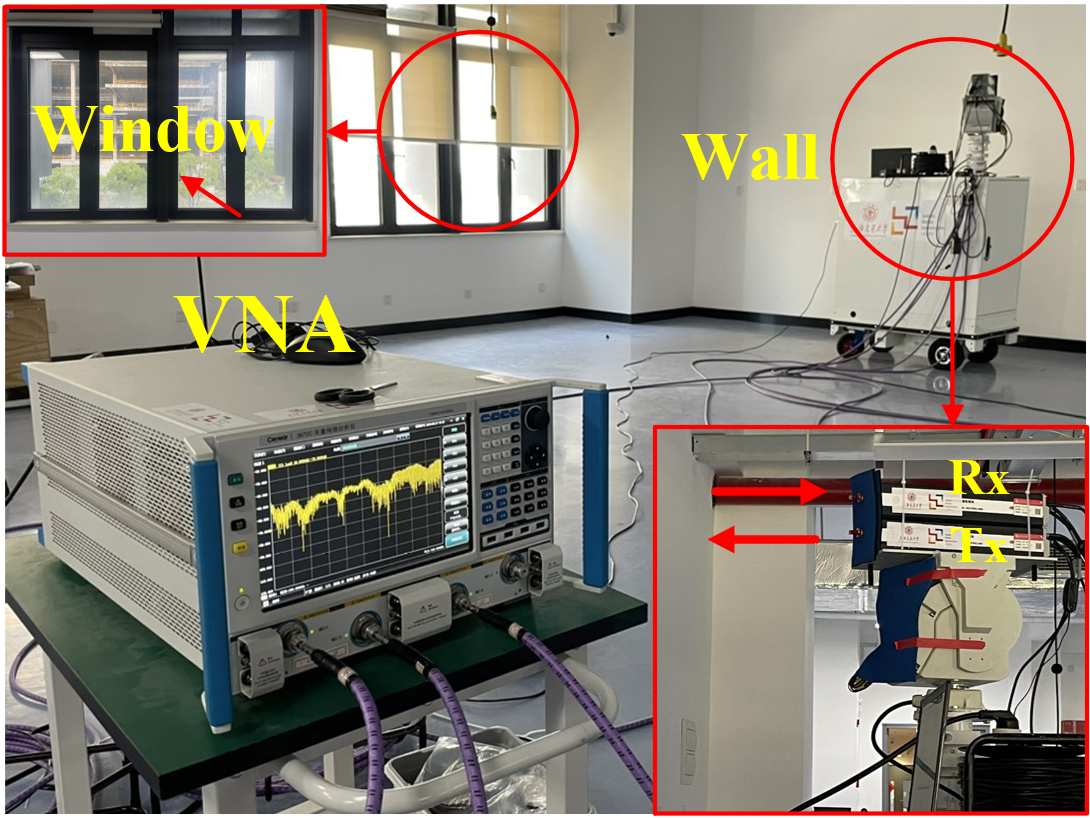}
\label{fig:scenario1}
}
\subfigure[]{
\includegraphics[width=0.36\textwidth]{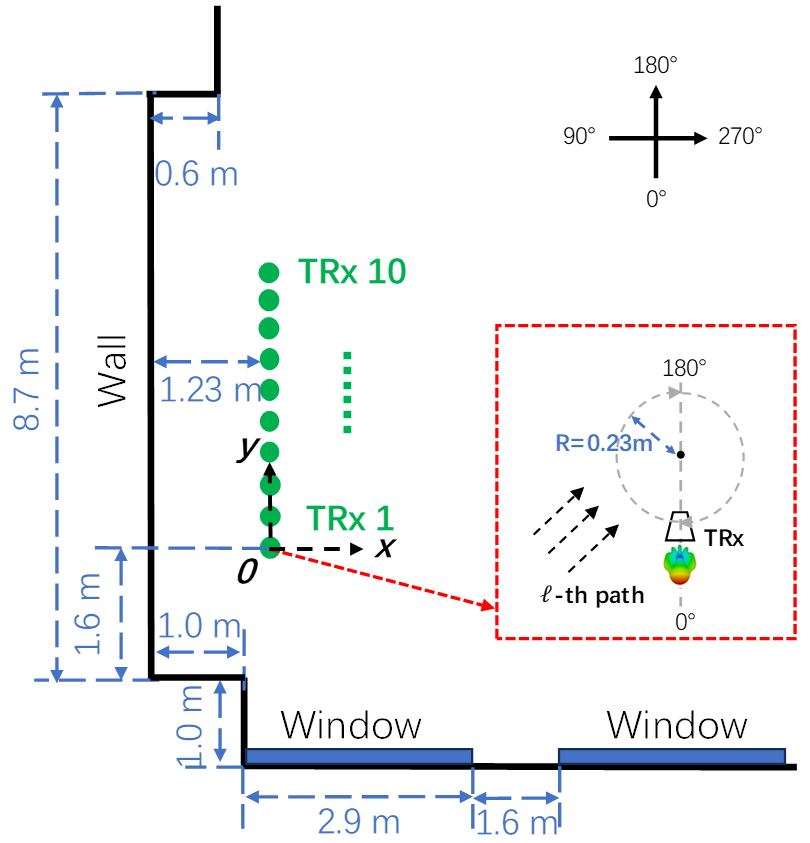}
\label{fig:scenario}
}
\captionsetup{font={footnotesize}}
\caption{Measurement system and scenario. (a) Real photo; (b) Schematic of the scenario.}
\label{fig:description}
\end{figure}

The monostatic sensing measurements are conducted in an empty area in a laboratory scenario, which is static and unoccupied, ensuring physical consistency of the environment, as shown in Fig.~\ref{fig:description}. In this measurement campaign, a VNA-based channel sounder is employed in~\cite{10494205,lyu2021design} with a frequency range of 290-310~GHz and 2001 frequency points. The directional scanning scheme (DSS) is applied to capture the channel spatial profiles. As shown in Fig.~\ref{fig:description}~(a), the transmitter (Tx) and receiver (Rx) are mounted on the same rotation platform, mimicking a transceiver (TRx), which investigates the THz wave interaction with the common indoor structures. Specifically, a uniform circular array (UCA) configuration is employed, in which the antennas are positioned in a circular layout with a radius $r$ of 0.23~m. The DSS is designed to provide uniform angular coverage, enabling high-resolution sensing across the azimuthal plane.

During the experiment, a 20~GHz wide band is measured, which corresponds to a delay resolution of 0.05~ns and a distance resolution of 1.5~cm. Note that the 1.5~cm distance resolution is determined from system level but the effective reconstruction accuracy can be further enhanced through advanced signal processing algorithms. The height of the monostatic TRx is 2.0~m. Moreover, both Tx/Rx are equipped with horn antennas, whose gain and half-power beamwidth (HPBW) are 26~dBi and 8$^{\circ}$, respectively. As depicted in Fig.~\ref{fig:description}~(b), to capture environment information from different directions, the TRx scans in the azimuth plane with 1$^{\circ}$ angle step, from 0$^{\circ}$ to 359$^{\circ}$. The measurements were conducted at 10 TRx locations, with 0.5~m distance separation between the adjacent TRx locations. While the TRx positions were automatically changed, resulting in variations in the scattering patterns due to changing relative observation angles, these variations enhanced the diversity of MPCs, thereby contributing to more comprehensive environmental information. The distance between each TRx location and the wall was uniformly maintained at 1.23~m. The experiment configurations are listed in Table~\ref{tab:Experiment parameters}. 

\begin{table}[]
\captionsetup{font={footnotesize}}
\caption{Experiment configurations}
\label{tab:Experiment parameters}
\centering
\begin{tabular}{cc}
\toprule
Parameter              & Value              \\
\midrule
Frequency band         & 290-310~GHz          \\
Bandwidth              & 20~GHz               \\
Delay resolution       & 0.05~ns             \\
Distance resolution & 1.5~cm               \\
TRx height             & 2.0~m                \\
Antenna gain of Tx/Rx  & 26~dBi               \\
HPBW of Tx/Rx antennas & 8$^{\circ}$                  \\
Azimuth rotation range & {[}0$^{\circ}$:1$^{\circ}$:359$^{\circ}${]} \\
\bottomrule
\end{tabular}
\end{table}

\section{CCA-assisted Channel Parameter Estimation}
In this section, the data processing procedures for environment reconstruction are illustrated. First, the measured channel frequency responses are transformed into the PADP to characterize multipath propagation. A CCA-assisted region segmentation method is then applied to extract high-power multipath regions. Furthermore, an element-wise SAGE algorithm is employed to estimate the delay and angle parameters of the dominant MPCs. 

\subsection{Signal Model}

In this work, the received signal model at a given rotation angle of UCA $\varphi$ and carrier frequency $f$ can be defined as
\begin{equation}
\label{equ:signal}
H(f, \varphi) = \sum_{\ell=1}^{L} \beta_{\ell} e^{-j 2 \pi f \tau_{\ell}},
\end{equation}
where $L$ is the total number of MPCs and $\tau_{\ell}$ denotes the propagation delay of the $\ell$-th MPC. The coefficient $\beta_{\ell}$ is the product of the amplitude $\alpha_{\ell}$ of the $\ell$-th path and the antenna array response, which can be expressed as
\begin{align}
\beta _{\ell}=\alpha _{\ell}\cdot a_{\mathrm{TRx},\ell}(f, \varphi).
\end{align}
where the antenna array response of TRx $a_{\mathrm{TRx},\ell}(f, \varphi)$ is given by
\begin{equation}
a_{\mathrm{TRx},\ell}(f, \varphi) =  e^{j 4 \pi f r \cos(\theta_{\ell} - \varphi)/\rm c} \cdot G_{\rm \mathrm{TRx}}(\theta_{\ell} - \varphi),
\end{equation}
where $r$, $\theta_{\ell}$, and $\rm c$ denote the radius of UCA, the angle of the $\ell$-th MPC, and the speed of light, respectively. The term $G_{\rm TRx}(\theta_{\ell} - \varphi)$ represents the summed antenna gain of the TRx at angle $\theta_{\ell}$. Each MPC can be characterized by the parameter set $\Theta = \{\alpha_{\ell}, \tau_{\ell}, \theta_{\ell}\}$. Note that since the Tx and Rx are co-located on the same rotation platform, the angle represents both the transmit and receive angles of the $\ell$-th MPC, i.e., $\theta_{\ell}=\theta_{Tx,\ell}=\theta_{Rx,\ell}$. 

\subsection{Region Segmentation}
In this work, a region segmentation method learnt from image processing is adopted to address the challenge of extracting dominant multipath regions from the PADP. By leveraging closing operation and connected component labeling (CCL) algorithm, the proposed approach segments high-energy regions in the PADP, enabling selective and computationally efficient parameter estimation for subsequent SAGE processing.

\begin{figure}
\centering
\includegraphics[width=0.45\textwidth]{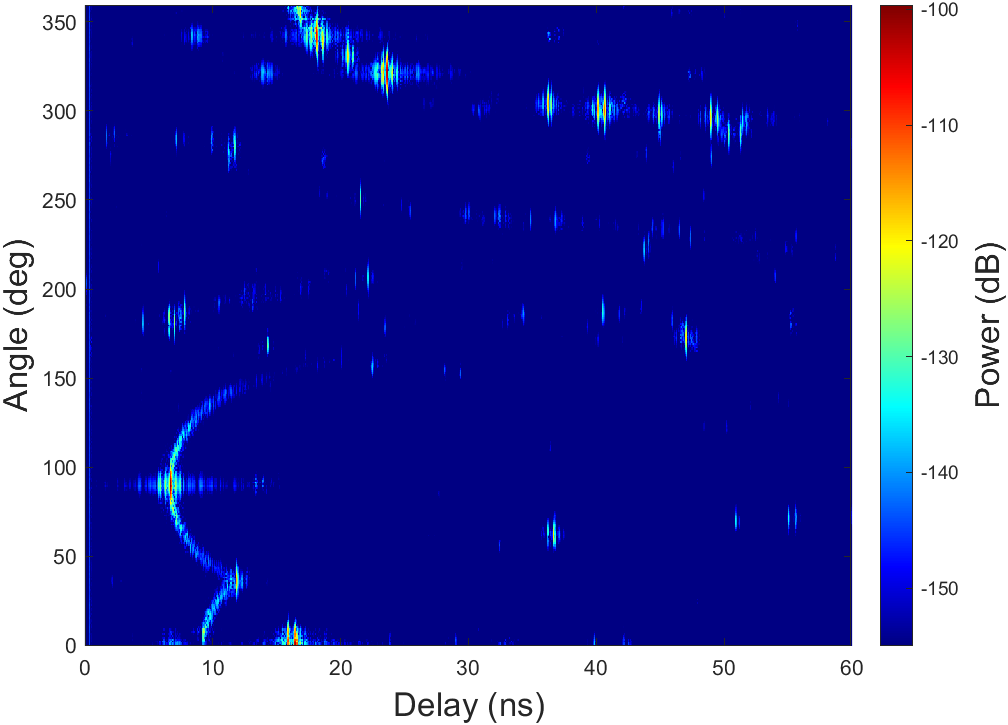}
\label{fig:scenario1}
\captionsetup{font={footnotesize}}
\caption{Exemplary PADP results at TRx 1 Location.}
\label{fig:PADP}
\end{figure}

The measurement results are obtained from VNA, i.e., the channel frequency response (CFR) ranging from 290~GHz to 310~GHz. The frequency-domain data is first calibrated and then transformed into the delay domain through the inverse discrete Fourier transform (IDFT) to obtain the channel impulse response (CIR). Specifically, the PADP is calculated by
\begin{equation}
P(\tau, \varphi) = 20\cdot \log_{10} \left( |h(\tau, \varphi)| \right),
\end{equation}
where $h(\tau,\varphi)$ denotes the received CIR, $\tau$ represents the propagation delay and $\varphi$ denotes the rotation angle of UCA. As shown in Fig.~\ref{fig:PADP}, the resulting PADP at TRx~1 Location provides the distribution of the MPCs in the delay and angle domain, capturing the interactions between the THz signal and the environment. A continuous high-power region is detected in the angle range of \(0^\circ\) to \(150^\circ\) and delays between \(5\) ns to \(13\) ns, suggesting the existence of a prominent consistent reflector. Additionally, several distinct regions with high power are visible in the range of \(280^\circ\) to \(360^\circ\), which may correspond to some materials with strong reflective property. 

Given the dense sampling in the delay and angular domains, applying the SAGE algorithm across the full delay and angle domain directly remains computationally infeasible, leading to prohibitive processing time and resource consumption. Therefore, a data-driven preprocessing strategy is necessary to constrain the search space before high-resolution parameter estimation. A CCA-assisted region search is applied to the PADP to enhance the efficiency of MPC estimation. The proposed method employs a CCA-based region search to extract high-power MPC regions, enabling selective parameter estimation within the most relevant multipath regions, thereby reducing the computational complexity of the subsequent SAGE-based MPC parameter estimation.

To identify high-power MPC regions, a thresholding operation is applied to remove low-power noise components and retain only the dominant reflections. A binary mask is constructed as

\begin{equation}
P_{\text{mask}}(\tau, \varphi) =
\begin{cases}
    1, & \text{if } P(\tau, \varphi) > P_{\text{th}} \\
    0, & \text{otherwise}
\end{cases},
\end{equation}
where \( P_{\text{th}} \) is a determined threshold set at 10~dB above the noise floor of the PADP~\cite{guan2020channel}.

To further refine the detected regions, morphological operations are employed to enhance region connectivity and eliminate small, spurious noise components. These operations include closing operation and CCL algorithm~\cite{gil2002efficient}, which iteratively improve the structure of the binary mask \( P_{\text{mask}}(\tau, \varphi) \) used for MPC extraction.

The closing operation is applied to fill small gaps in detected regions and ensure more robust region extraction, which can be expressed as
\begin{equation}
P_{\text{closed}}(\tau, \varphi) = (P_{\text{mask}} \oplus S) \ominus S,
\end{equation}
where $\oplus$, $\ominus$, and $S$ denote the morphological dilation operator, erosion operator, and structuring element, respectively. The output $P_{\text{closed}}(\tau, \varphi)$ serves as the optimized binary mask for subsequent region segmentation operation.

\begin{figure}[t]
\centering
\includegraphics[width=0.45\textwidth]{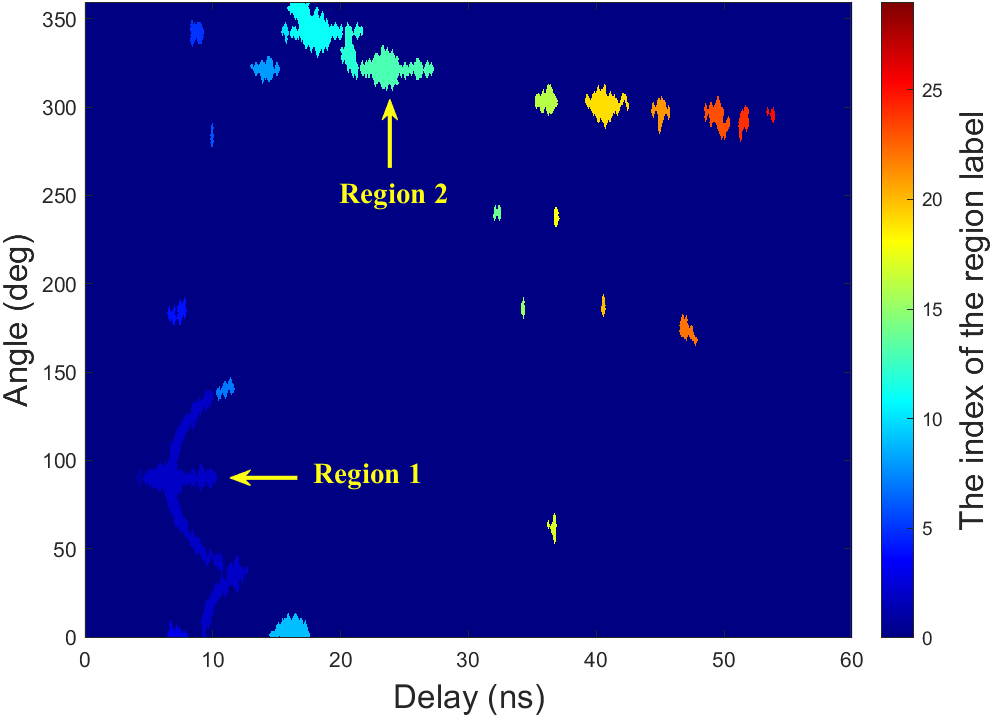}
\captionsetup{font={footnotesize}}
\caption{CCA result for the PADP at TRx 1 Location.}
\label{fig:PADP_CV}
\end{figure}

Then a CCL algorithm is applied to segment contiguous high-energy MPC regions using an 8-connectivity rule, which considers a pixel as connected if it shares either an edge or a corner with another pixel in the neighborhood~\cite{haralick1985image,jahne2005digital}. This ensures that diagonal connections between adjacent high-power regions are preserved, leading to more accurate segmentation of multipath components.

The labeling process is defined as follows:
\begin{equation}
I(\tau, \varphi) = \sum_{(m,n) \in \mathcal{N}} \mathbb{I} (P_{\text{closed}}(m, n) = 1),
\end{equation}
where \( \mathcal{N} \) defines the neighborhood structure and \( \mathbb{I}(\cdot) \) is an indicator function that identifies pixels belonging to the same component.

Each connected component corresponds to a distinct multipath region, within which all MPCs are assigned the same label \( I(\tau, \varphi) \), enabling localized and region-wise parameter estimation in the subsequent SAGE processing. Components smaller than a predefined minimum pixel count \( N_{\text{min}} \) are discarded to eliminate noise~\cite{he2017connected}. The valid high-power multipath regions are stored in a binary mask \( m(i, j) \), which defines the searchable region for SAGE estimation. 

For each identified connected component \( L_i \), the set of associated delay and angle values is extracted as:

\begin{equation}
\mathcal{R}_i = \{ (\tau, \theta) \ | \ I(\tau, \varphi) = L_i, \ m(i, j) = 1 \},
\end{equation}
where \( I(\tau, \varphi) \) is the labeling function that assigns a unique identifier to each connected component, and \( m(i, j) = 1 \) ensures that only the high-power multipath regions are considered. \( i \) and \( j \) correspond to the discrete indices of the delay and angle dimensions in the PADP matrix:
\begin{equation}  
i = \arg \min_{i} |\tau_i - \tau|, 
\end{equation}
\begin{equation}  
j = \arg \min_{j} |\theta_j - \theta|.
\end{equation}

The extracted multipath regions are then passed to the SAGE algorithm, which iterates over all identified regions \( L_i \):

\begin{equation}
\mathcal{R} = \bigcup_{i} \mathcal{R}_i.
\end{equation}

Instead of performing exhaustive parameter search across the entire PADP, the proposed method restricts the SAGE search space to the segmented multipath regions identified by the CCA. By iterating over all \( L_i \), the SAGE algorithm estimates the dominant multipath components within each region, thereby reducing computational complexity while preserving estimation accuracy.
The processed multipath regions after CCA-assisted segmentation are shown in Fig.~\ref{fig:PADP_CV}, where each connected component is assigned a unique label. Different colors represent distinct multipath regions, refining the spatial structure information. 

\begin{figure*}[t]
\centering
\includegraphics[width=1.95\columnwidth]{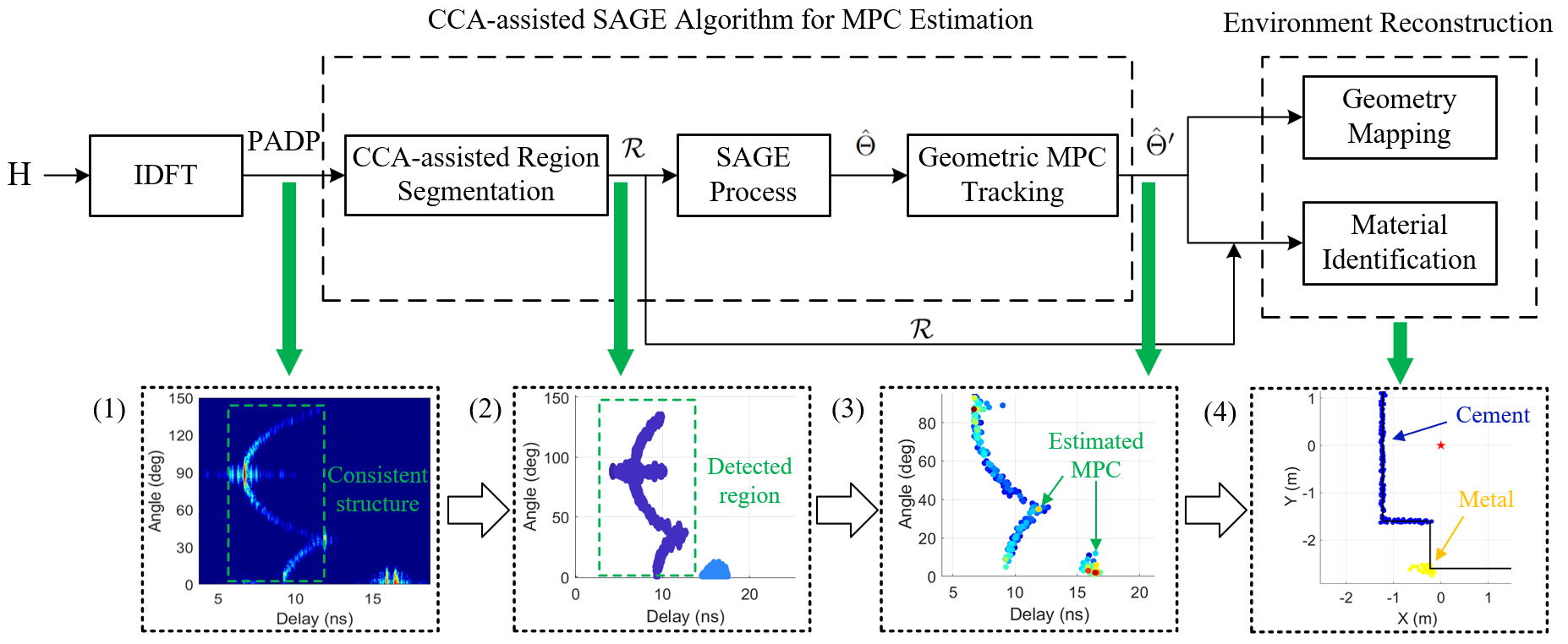}
\captionsetup{font={footnotesize}}
\caption{CCA-assisted SAGE Algorithm for THz-Based Reconstruction.}
\label{fig:flowchart}
\end{figure*}

\subsection{MPC Parameters Estimation}

To estimate MPCs in the environment, we adopt an MPC-tracking-based SAGE algorithm to estimate the delay and angle of MPCs, readers can refer to~\cite{mpc_estimation} for details of the algorithm. The MPC-tracking-based SAGE algorithm has two stages, i.e., element-wise SAGE estimation and geometry-based MPC tracking. In the element-wise SAGE estimation, the SAGE algorithm is applied to the CFR of each rotation angle to estimate the amplitude and delay of each MPC. In this stage, the estimated parameter set can be expressed as $\varTheta^{(k)} = [\varTheta^{(k)}_{1}, \varTheta^{(k)}_{2}, \cdots, \varTheta^{(k)}_{359}]$, where $k$ denotes the index of the TRx location. For each $\varTheta^{(k)}_{\varphi}$, the matrix contains the parameters for the MPCs in the rotation angle of $\varphi$ with a total MPC number of $L$, which can be expressed as 
 \begin{align}
\hat{\varTheta}^{(k)}_{\varphi}=\left[ \begin{matrix}
\hat{\beta} _{1}^{(k,\varphi)}&	\cdots &    \hat{\beta} _{L}^{(k,\varphi)}\\
\hat{\tau} _{1}^{(k,\varphi)}&    \cdots &    \hat{\tau} _{L}^{(k,\varphi)}\\
\end{matrix} \right],
 \end{align}

By incorporating the CCA-assisted multipath regions as priors and considering the antenna radiation pattern from the measurement, the element-wise SAGE algorithm achieves improved computational efficiency while allowing for high resolution in environment reconstruction. 

After obtaining the estimated MPC parameters, the geometry-based MPC trajectory tracking can be employed to classify the MPC with the same antenna pattern effect \cite{mpc_estimation}. We can then simply take the MPCs with the maximum power in each MPC trajectory as the de-embedded MPC. The de-embedded parameter set $\hat{\varTheta}^{(m)}_{\rm de}$ can be expressed as
 \begin{align}
\hat{\varTheta}^{(k)}_{\rm de}=\left[ \begin{matrix}
\hat{\alpha}_{1}^{(k)}&	\cdots & \hat{\alpha}_{\ell}^{(k)} &\cdots &  \hat{\alpha}_{L}^{(k)}\\
\hat{\tau}_{1}^{(k)}&    \cdots & \hat{\tau}_{\ell}^{(k)} &\cdots &    \hat{\tau}_{L}^{(k)}\\
\hat{\phi}_{1}^{(k)}&    \cdots & \hat{\phi}_{\ell}^{(k)} &\cdots &    \hat{\phi}_{L}^{(k)}\\
\end{matrix} \right].
 \end{align}

The overall processing flow of the proposed CCA-assisted SAGE algorithm is illustrated in Fig.~\ref{fig:flowchart}. The raw measured channel frequency response $H$ is first transformed into the PADP through IDFT, as shown in subfigure~(1). A CCA-assisted region segmentation is then applied to the PADP to identify high-power multipath regions. At this step, a binary segmentation mask is generated to indicate valid MPC regions, and each identified region is assigned a unique label, as visualized in subfigure~(2). The resulting set of segmented regions $\mathcal{R}$ serves as the search space for the subsequent SAGE-based MPC parameter estimation. Element-wise SAGE is then performed within each labeled region to estimate the delay and angle of multipath components efficiently. The estimated MPCs are further organized using geometric MPC tracking, as shown in subfigure~(3). Finally, the estimated MPCs are then utilized for overall environment reconstruction. The geometry mapping module leverages the estimated delay and angle information to reconstruct the environmental layout, while the material identification module analyzes the reflection characteristics of detected regions to identify materials, as illustrated in subfigure~(4).

\section{Methodology for Environment Reconstruction}

In this section, the methodologies for geometry mapping and material identification are presented, enabling spatial reconstruction and material sensing based on the estimated MPC parameters.

\subsection{Geometric Mapping}
\label{sec:mapping}
\begin{figure*}
\centering
\subfigure[]{
\includegraphics[width=0.23\textwidth]{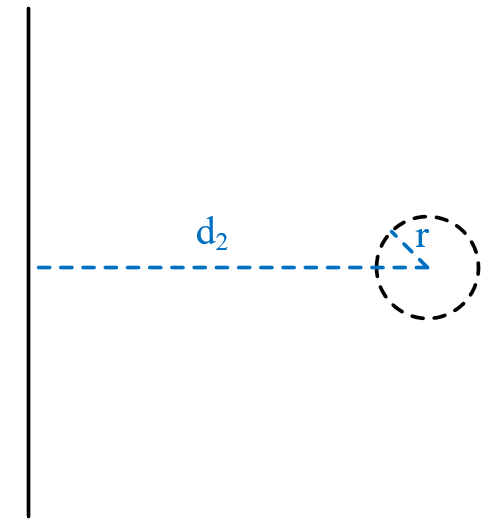}
\label{fig:ccaa}
}
\subfigure[]{
\includegraphics[width=0.23\textwidth]{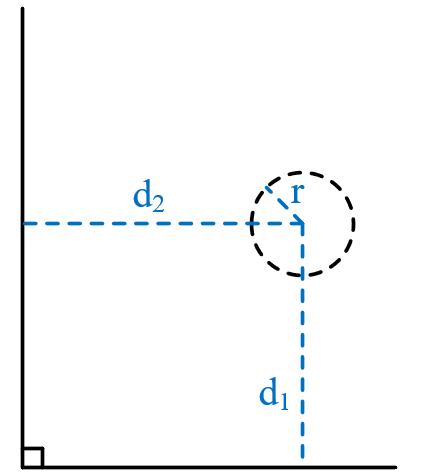}
\label{fig:ccab}
}
\subfigure[]{
\includegraphics[width=0.23\textwidth]{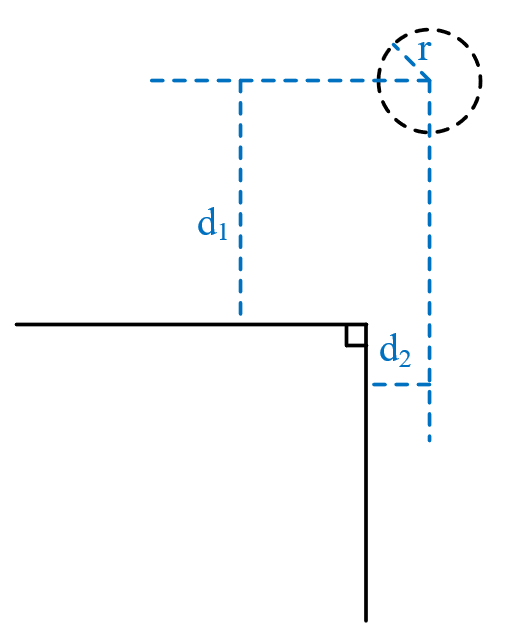}
\label{fig:ccac}
}
\subfigure[]{
\includegraphics[width=0.27\textwidth]{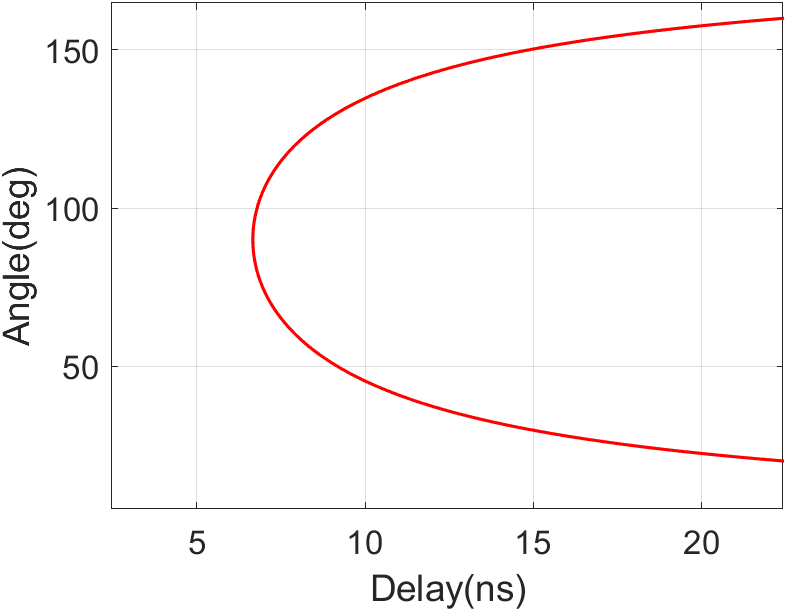}
\label{fig:ccad}
}
\subfigure[]{
\includegraphics[width=0.27\textwidth]{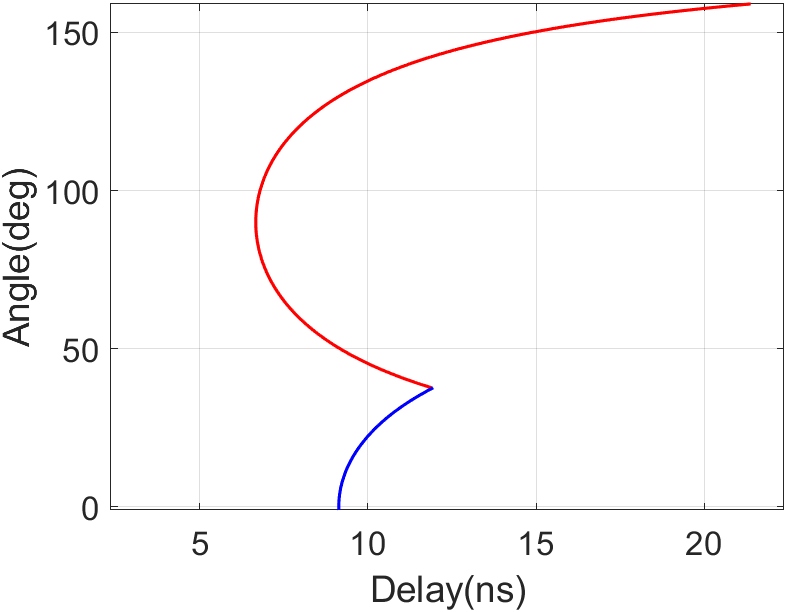}
\label{fig:ccae}
}
\subfigure[]{
\includegraphics[width=0.27\textwidth]{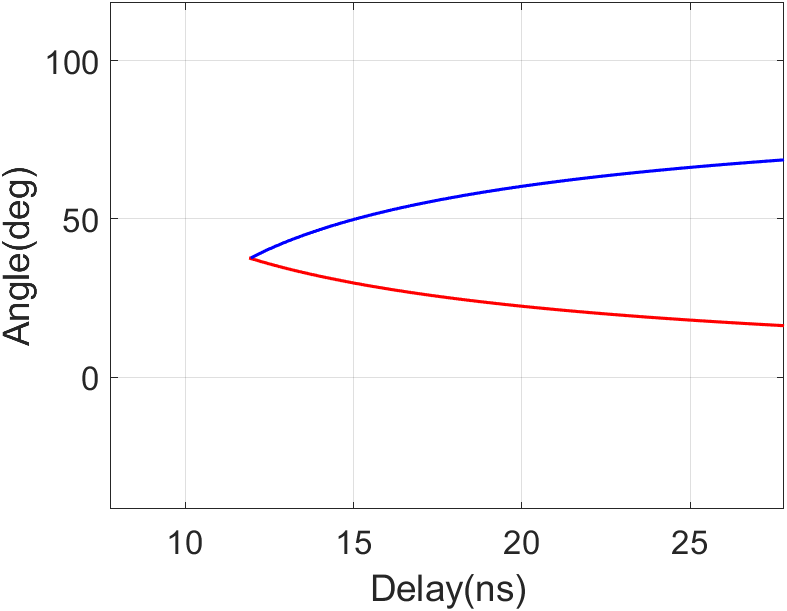}
\label{fig:ccaf}
}
\captionsetup{font={footnotesize}}
\caption{Theoretical PADP results under typical indoor structures. (a) Schematic diagram of a flat wall; (b) Schematic diagram of an inner corner; (c) Schematic diagram of an outer corner; (d)-(f) Corresponding PADP from single-bounce reflections for the three cases.}
\label{fig:CCA}
\end{figure*}

After estimating the parameters of each MPC, specifically the delay and angle, the detected components can be mapped into the geometric domain. For details of the mapping equations, readers are kindly referred to~\cite{fang2024centimeter}. In indoor environments, walls and corners are among the most prevalent structural features. Therefore, in this work, we further model and analyze the structure of the connected component regions under such conditions. Theoretical delay-angle models are developed based on typical indoor geometries. Specifically, three common structures are considered: a flat wall, an inner corner, and an outer corner, as illustrated in Fig.~\ref{fig:CCA}~(a)-(c). Taking the inner-corner structure as an example, the delay $\tau$ can be analytically expressed as a function of the rotation angle $\theta$ under the assumption of single-bounce reflection:
\begin{equation}
    \tau_{\text{left}}(\theta) = \frac{2}{c}\left(\frac{d_1}{\sin \theta} - r\right), 
\end{equation}
\begin{equation}
    \tau_{\text{bottom}}(\theta) = \frac{2}{c}\left(\frac{d_2}{\cos \theta} - r\right),
\end{equation}
where $d_1$ and $d_2$ denote the perpendicular distances from the TRx center to the left and bottom walls, respectively, $r$ is the rotation radius of the antenna array, and $c$ is the speed of light. The resulting theoretical PADP patterns are shown in Fig.~\ref{fig:CCA} (d)-(f), which serve as templates for structure inference from connected components. To further extract the structural parameters of the connected component region, we formulate an optimization problem to minimize the RMSE between the measured delays and theoretical predictions:
\begin{equation}
    \min_{d_1, d_2} \sqrt{\frac{1}{N} \sum_{i=1}^N \left( \tau_i^{\text{meas}} - \tau_i^{\text{model}}(d_1, d_2, \theta_i) \right)^2}, \label{eq:rmse}
\end{equation}
where $(\theta_i, \tau_i^{\text{meas}})$ denotes the $i$-th MPC in the region, and $\tau_i^{\text{model}}$ is the corresponding theoretical delay under the inner-corner assumption.

After the connected component structure is identified, a sliding-window-based filtering technique is applied to further improve the mapping accuracy and suppress residual noise in the geometry mapping, which smooths out local fluctuations in the estimated positions. Given the set of mapping points \(\mathcal{M} = \{(x_i, y_i)\}\), a moving average filter is applied to obtain smoothed coordinates:
\begin{equation}
    x_i^{\text{fil}} = \frac{1}{W} \sum_{j=i-W/2}^{i+W/2} x_j,
\end{equation}
\begin{equation}
    y_i^{\text{fil}} = \frac{1}{W} \sum_{j=i-W/2}^{i+W/2} y_j,
\end{equation}
where \( W \) is the window size, and \( (x_i^{\text{fil}}, y_i^{\text{fil}}) \) represents the refined position of each mapping point.

% The optimal window size \( W_{\text{opt}} \) is determined by minimizing the MDE between the mapping points and the ground truth:
% \begin{equation}
% \text{MDE}(W) = \frac{1}{N} \sum_{i=1}^{N} \left( | x_i(W) - x_{\text{ref}} | + | y_i(W) - y_{\text{ref}} | \right),
% \end{equation}
% where \( (x_{\text{ref}}, y_{\text{ref}}) \) represents the reference boundary of the environment. 

By leveraging this adaptive filtering approach, the proposed method refines the mapping structure while effectively mitigating the distortions introduced by noise and multipath interference, thereby enhancing the overall accuracy of THz-based geometry mapping.

\subsection{Material Identification}
% \begin{figure}[t]
% \centering
% \subfloat[Simplified schematic diagram of the THz-TDS.]{\includegraphics[width=0.8\columnwidth]{figures/TDS2.png}}%
% \label{fig:tds1}
% \hfil
% \subfloat[A real photo for the material measurements.]{\includegraphics[width=0.6\columnwidth]{figures/TDS.png}}%
% \label{fig:tds2}
% \caption{THz-TDS system for material measurements.}
% \label{fig:tds}
% \end{figure}

\begin{figure}
\centering
\subfigure[]{
\includegraphics[width=0.4\textwidth]{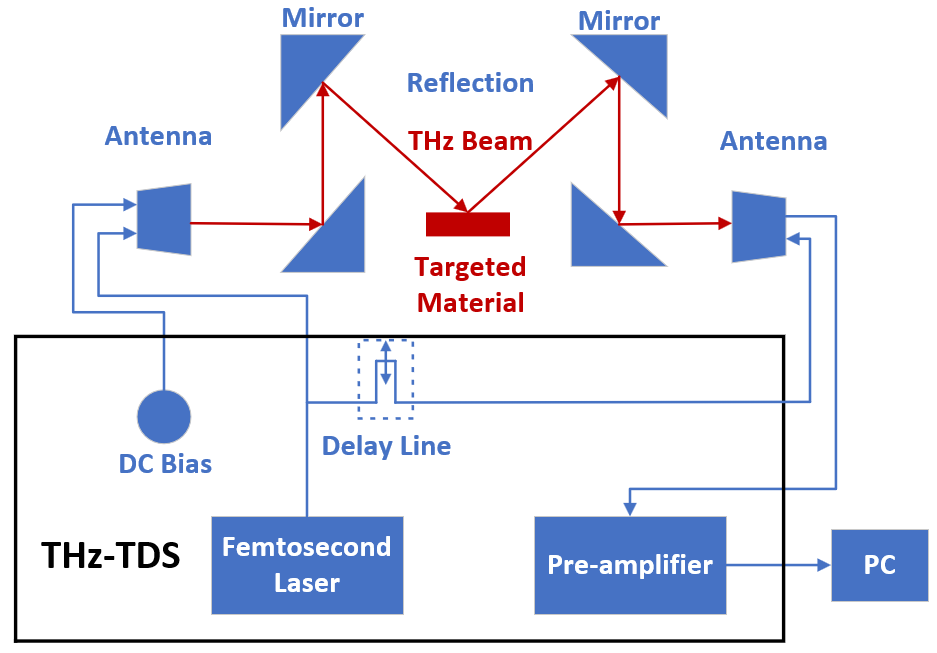}
\label{fig:tds1}
}
\subfigure[]{
\includegraphics[width=0.3\textwidth]{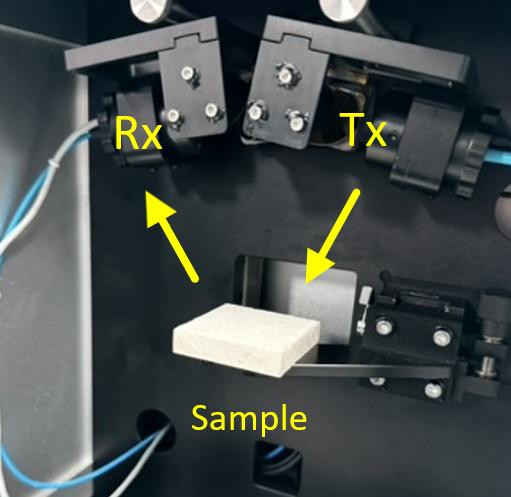}
\label{fig:tds2}
}
\captionsetup{font={footnotesize}}
\caption{THz-TDS system for material measurements. (a) Simplified schematic diagram of the THz-TDS; (b) A real photo for the material measurements. }
\label{fig:tds}
\end{figure}

Furthermore, we cross-reference the material measurements with the channel measurement results. As illustrated in Fig.~\ref{fig:tds}, THz-TDS is utilized to characterize the reflection properties of various materials. The THz-TDS system covers a broad frequency range of 0–6~THz and employs ultrafast electromagnetic pulses in the THz spectrum to probe the material samples~\cite{lyu2023sub}. A short laser pulse generates a broadband THz pulse, which interacts with the material under test. The transmitted THz pulse is then detected by a photoconductive antenna, providing time-domain data for further analysis. To accurately extract the reflection properties of the materials, measurements are conducted with a $20^{\circ}$ angle between the Tx and Rx, corresponding to an incidence angle of $10^{\circ}$. To ensure measurement accuracy, we perform a reference calibration using a gold mirror, which serves as an ideal reflector with negligible loss. The material response for each target sample is then normalized against this reference measurement, effectively mitigating the influence of system response variations. The sample under test is positioned at a fixed 9~cm from the antennas. 

The material reflection database is presented based on the measurements through THz-TDS, which initially captures the material response in the delay domain, and then is converted into the frequency domain using a discrete Fourier transform (DFT) function across 0-6~THz frequency range. Note that the free-space path loss (FSPL) is negligible due to the short distance of 9~cm. In order to be consistent with the channel measurement, the measured reflection loss values $RL$ are recorded at 300~GHz. Note that $RL$ for each material sample is measured at an incidence angle $\gamma_1 = 10^\circ$ in the THz-TDS setup, where $\gamma_1$ denotes the angle between the incident THz wave and the sample surface normal. To ensure consistency with normal-incidence channel measurements, the measured value is calibrated to the equivalent normal-incidence condition, i.e., $\gamma_1 = 0^\circ$, using the Fresnel equations and Snell’s law~\cite{series2015effects}.

The reflection coefficient $R$ is calculated as
\begin{equation}
R = \frac{\sqrt{\eta_1} \cos \gamma_1 - \sqrt{\eta_2} \cos \gamma_2}
         {\sqrt{\eta_1} \cos \gamma_1 + \sqrt{\eta_2} \cos \gamma_2},
\end{equation}
where $\gamma_2$ is the transmission angle. $\eta_1$ and $\eta_2$ denote the complex relative permittivities of air and the sample material, respectively.  The $\cos \gamma_2$ term can be evaluated in terms of $\gamma_1$ using

\begin{equation}
\cos\gamma_2 = \sqrt{1 -  \frac{\eta_1}{\eta_2} \sin^2\gamma_1  }.
\end{equation}

The reflection loss in decibels (dB) is then computed as
\begin{equation}
RL(\gamma_1) = -10 \log_{10} R.
\end{equation}

In our case, assuming an air permitivity $\eta_1$ of 1 F/m, we first obtain $RL(10^\circ)$ from the THz-TDS measurement, and use this value to calculate the permittivity $\eta_2$ of the material. With the known $\eta_2$, we then calculate the theoretical reflection loss $RL(0^\circ)$ for normal incidence, and record this value in the material database. This procedure ensures that all database entries are referenced to normal incidence, providing consistency and comparability with the channel measurement configuration. To be consistent with the channel measurement, the measured reflection loss values are recorded at 300~GHz.

The experiments are conducted with a wide variety of common materials, which are categorized into biological materials, metals, building materials, and functional materials. Under each major category, a set of material types is defined according to their specific physical and electromagnetic characteristics. For each material type, multiple samples are measured and the reflection loss values are averaged to ensure statistical reliability. The total test set includes over 200 material samples and is classified into more than 30 types.

Based on the channel measurement data, CCA-assisted SAGE algorithm for MPC estimation is used to extract the dominant regions. The reflection loss of each identified region is computed by

\begin{equation} 
RL_\ell^{q} = P_{\text{loss}, \ell}^{q} - \text{FSPL}_\ell^{q},
\end{equation}
where $RL_\ell^{q}$, $P_{\text{loss}, \ell}^{q}$ and $\text{FSPL}_\ell^{q}$ represent the reflection loss, estimated power and FSPL of $\ell$-th MPC in region $q$, respectively.

The FSPL of each MPC is given by
\begin{equation} 
\text{FSPL}_\ell = 20 \cdot \log{10}(4 \pi f \tau_\ell). 
\end{equation}

For each identified label (i.e., MPC region) in the segmented PADP, reflection losses are calculated for all MPCs within that region. In monostatic sensing scenarios, the dominant propagation mechanisms are specular and diffuse reflections, allowing the reflection behavior to be analyzed accordingly. Specular reflections, representing the deterministic component of the channel, are characterized by strong and concentrated energy at a single delay-angle pair. In this case, the minimum reflection loss within the region is used for material identification. In contrast, diffuse reflections represent the stochastic component, where the reflected MPCs exhibit a spread of power across multiple delays and angles, indicating scattering effects caused by rough surfaces. Based on this principle, a threshold can be chosen for each region to distinguish between deterministic and stochastic components. Low reflection loss values within the region can be selected and compared against the THz-TDS material database to identify the most likely material match.

\section{Results Analysis}
\begin{figure}[t]
\centering
\includegraphics[width=0.45\textwidth]{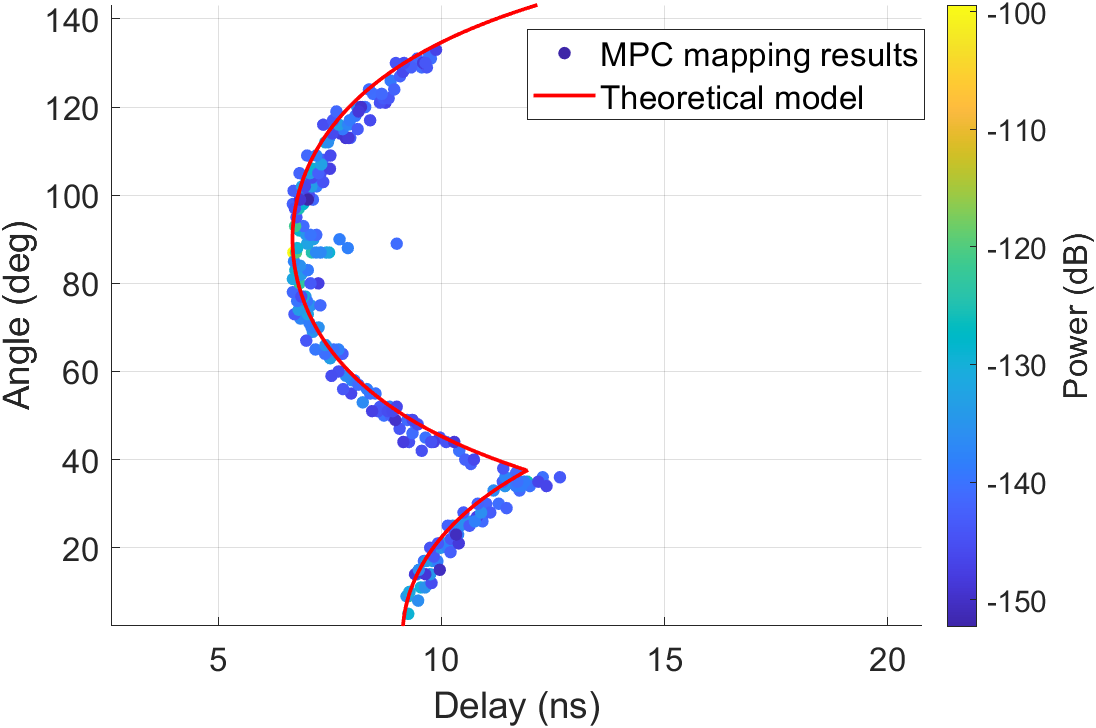}
\captionsetup{font={footnotesize}}
\caption{Comparison between estimated PADP and the theoretical model.}
\label{fig:CCA_comparison}
\end{figure}

\begin{figure}[t]
\centering
\includegraphics[width=0.40\textwidth]{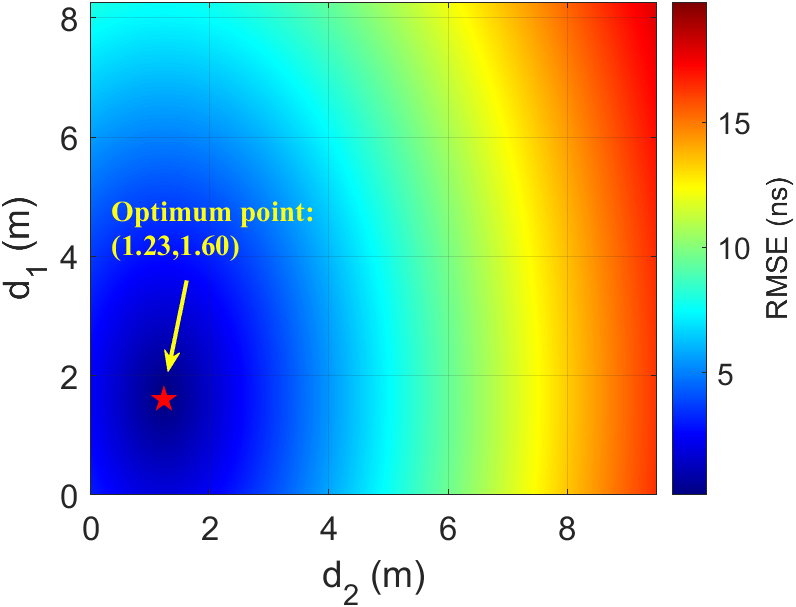}
\captionsetup{font={footnotesize}}
\caption{Optimization surface for delay-angle model fitting based on estimated MPCs.}
\label{fig:optimization}
\end{figure}

In this section, an analysis of the CCA-assisted channel parameter estimation algorithm is presented, along with a comparison against conventional algorithms. We also demonstrate geometry mapping results based on channel measurements, achieving millimeter-level accuracy. In addition, material identification results are provided based on the comparison between measured channel data and the material database.

\begin{table*}[t]
\caption{Comparison of computational complexity for MPC estimation per measurement location.}\label{tab:complexity}
\label{tab:complexity_vertical}
\centering
\begin{tabular}{|c|c|c|c|}
\hline
Algorithm & 2D-SAGE in~\cite{fang2024centimeter} & Element-wise SAGE in~\cite{mpc_estimation} & Proposed CCA-assisted SAGE\\
\hline
Search space & \multicolumn{3}{c|}{$0-360^\circ$ in angular domain, $0-60$~ns in delay domain} \\
\hline
Search resolution & \multicolumn{3}{c|}{1$^\circ$ in angular domain, 0.05~ns in delay domain} \\
\hline
Computational time & 72~h & 42~min & 5~min \\
\hline
\end{tabular}
\end{table*}

\subsection{Computational Complexity}
The computational complexity of MPC estimation is a critical factor in achieving high-resolution THz-based sensing. In this study, the computational efficiency of the proposed CCA-assisted SAGE algorithm is compared with the previous 2D-SAGE method under identical parameter resolution settings. As described in~\cite{fang2024centimeter}, the conventional 2D-SAGE algorithm performs exhaustive grid searches over a predefined parameter space, resulting in significant computational overhead over the full parameter space, especially when high resolution is required.

To ensure a fair comparison, both methods are evaluated under a search space of \([0,360^\circ]\) for AoA and \([0,60]\)~ns for delay, with a fixed resolution of \(1^\circ\) and 0.05~ns in the angle and delay domain, respectively. Note that the number of iterations is set to 1. As summarized in Table~\ref{tab:complexity}, the conventional 2D-SAGE algorithm requires a substantial computation time of 72~h for a single TRx location. In contrast, the element-wise SAGE algorithm proposed in~\cite{mpc_estimation} significantly reduces the time to 42~min. Building on this, the proposed CCA-assisted SAGE algorithm further accelerates the computation time to just 5~min per location, an 8.4$\times$ improvement, while maintaining the same estimation accuracy. This notable speedup is achieved through a selective parameter estimation strategy enabled by CCA-based region segmentation, which eliminates redundant computations in low-energy regions of the PADP. Additionally, our implementation supports an enhanced delay resolution of 0.01~ns, highlighting the scalability and efficiency of the proposed method for high-resolution ISAC applications.

\subsection{Geometry Mapping}
\begin{figure}
\centering
\includegraphics[width=1\columnwidth]{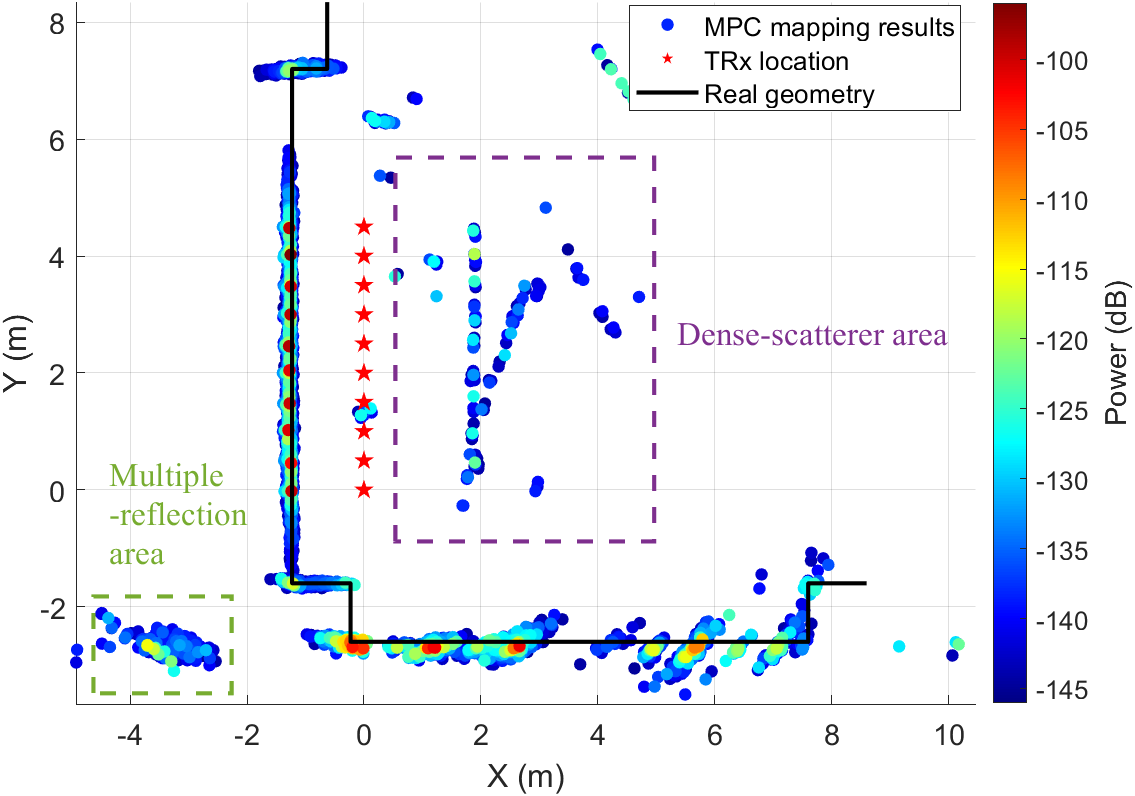}
\captionsetup{font={footnotesize}}
\caption{Overall geometry mapping result from multiple TRx locations using the proposed algorithm.}
\label{fig:overall_geometry}
\end{figure}

\begin{figure}
\centering
\subfigure[]{
\includegraphics[width=0.43\textwidth]{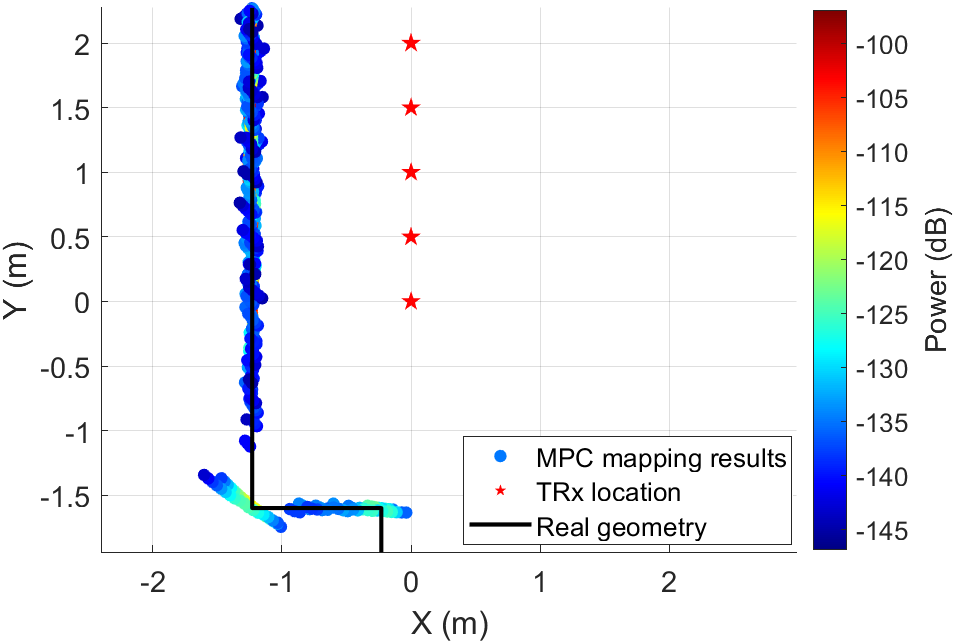}
\label{fig:recon_maxp}
}
\subfigure[]{
\includegraphics[width=0.43\textwidth]{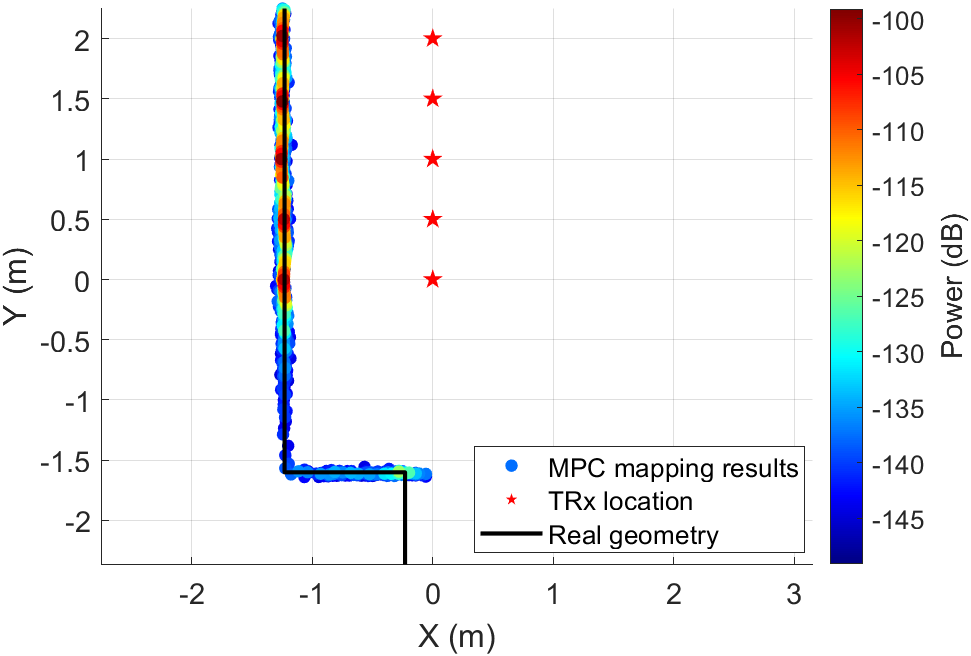}
\label{fig:recon_MLE}
}
\subfigure[]{
\includegraphics[width=0.43\textwidth]{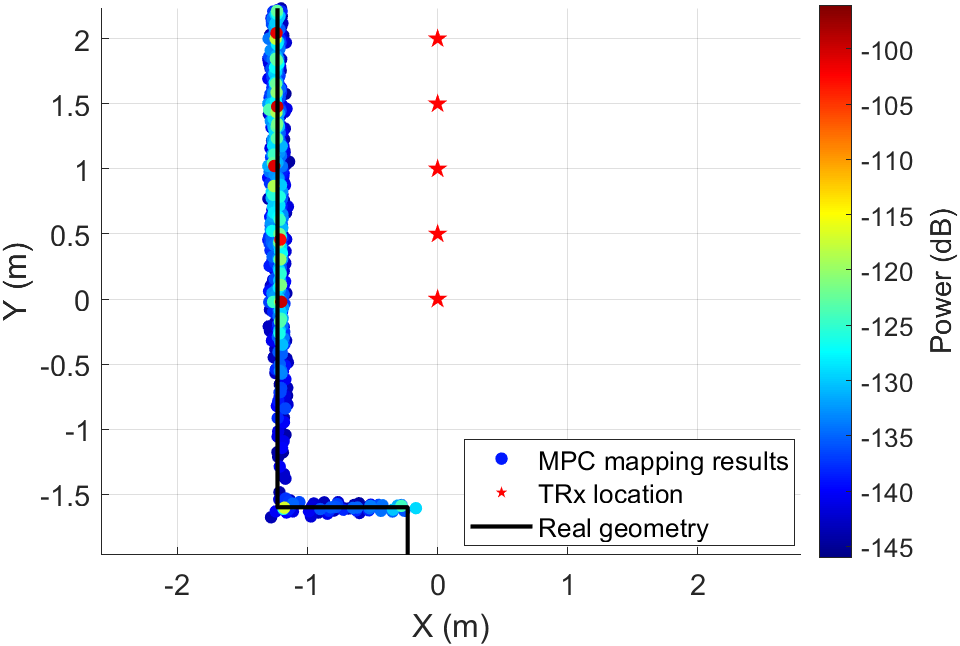}
\label{fig:recon_corner}
}
\captionsetup{font={footnotesize}}
\caption{Geometry mapping results based on conventional algorithms. (a) Maximum-search algorithm; (b) 2D SAGE algorithm; (c) Previous algorithm in \cite{mpc_estimation}.}
\label{fig:recon_envir}
\end{figure}
From the CCA-assisted SAGE algorithm results, we can detect the connect component regions. For example, as shown in Fig.~\ref{fig:PADP_CV}, 29 dominant regions can be detected. In this section, for geometry mapping, the first step is to identify the structure of the detected regions.
The extracted CCA results are further analyzed, and their structural parameters (i.e., geometric descriptors such as wall distance, intersection point, and corner angle) are fitted using the method proposed in Section~IV-A. Taking one connected component extracted from TRx~1 as an example, this region exhibits a characteristic two-branch structure in the delay-angle domain, indicative of corner-induced MPC, which matches the inner-corner geometry closely among the theoretical models. As illustrated in Fig.~\ref{fig:CCA_comparison}, the estimated MPCs are aligned well with the theoretical model. We use (\ref{eq:rmse}) to fit the best structural parameters, i.e., $d_1$ and $d_2$, and the RMSE surface is shown in Fig.~\ref{fig:optimization}. The optimal result occurs at $d_1 = 1.60$~m and $d_2 = 1.23$~m, achieving a minimum RMSE of $0.37$~ns, which is in close agreement with the actual TRx-wall distances measured in the real scenario. By identifying the structure of the targeted regions, the sliding-window filtering technique can be applied to reduce the geometry error.

% \begin{figure}
% \centering
% \subfigure[]{
% \includegraphics[width=1\columnwidth]{figures/recon_total.png}
% \label{fig:all_sage}
% }
% \subfigure[]{
% \includegraphics[width=1\columnwidth]{figures/recon_MLE_all.png}
% \label{fig:all_mle}
% }
% \captionsetup{font={footnotesize}}
% \caption{Overall geometry reconstruction result from multiple TRx locations. (a) 2D SAGE algorithm; (b) Proposed algorithm.}
% \label{fig:overall_geometry}
% \end{figure}

By calculating the geometric positions of the estimated MPCs, the reconstructed environment is visualized in Fig.~\ref{fig:overall_geometry}, which represents the aggregated mapping results from 10 TRx locations. A clear distinction can be observed between the well-defined structures, such as the walls and the scattered MPCs. Several dense linear regions are observed around structural edges, indicating strong specular reflections from flat surfaces. Additionally, multiple reflections from the dense-scatterer area, i.e., surrounding equipment and laboratory objects, contribute to outlier points in other directions. Nevertheless, the major reflective surfaces, particularly the wall, are accurately reconstructed. To quantitatively evaluate the reconstruction accuracy, we compute the geometry error based on the wall structure.

\begin{figure}
\centering
\includegraphics[width=0.45\textwidth]{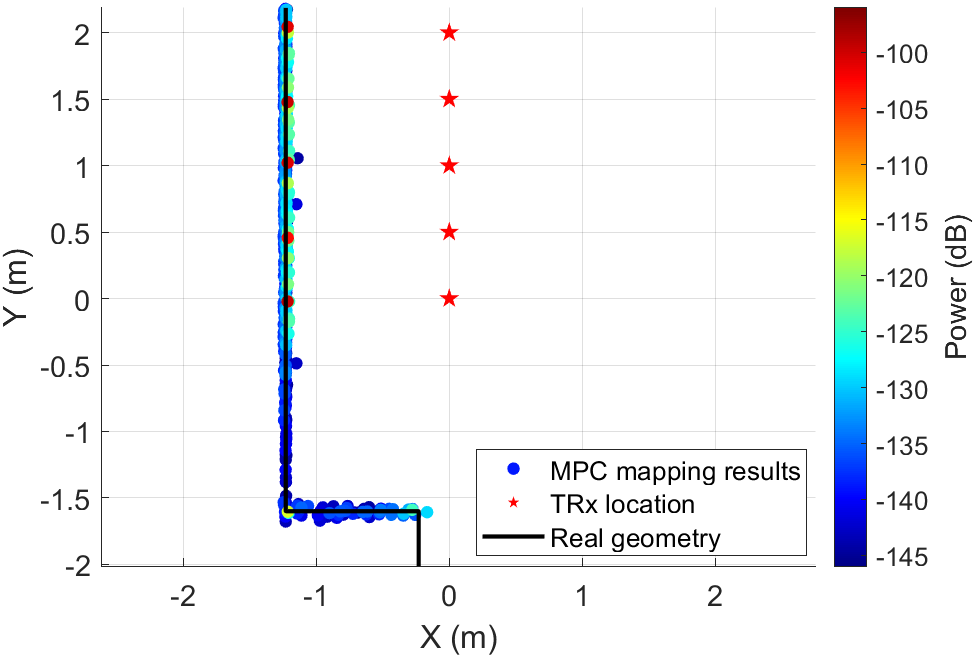}
\captionsetup{font={footnotesize}}
\caption{Wall mapping results based on proposed algorithm.}
\label{fig:recon_new}
\end{figure}

\begin{figure}[t]
\centering
\includegraphics[width=0.9\columnwidth]{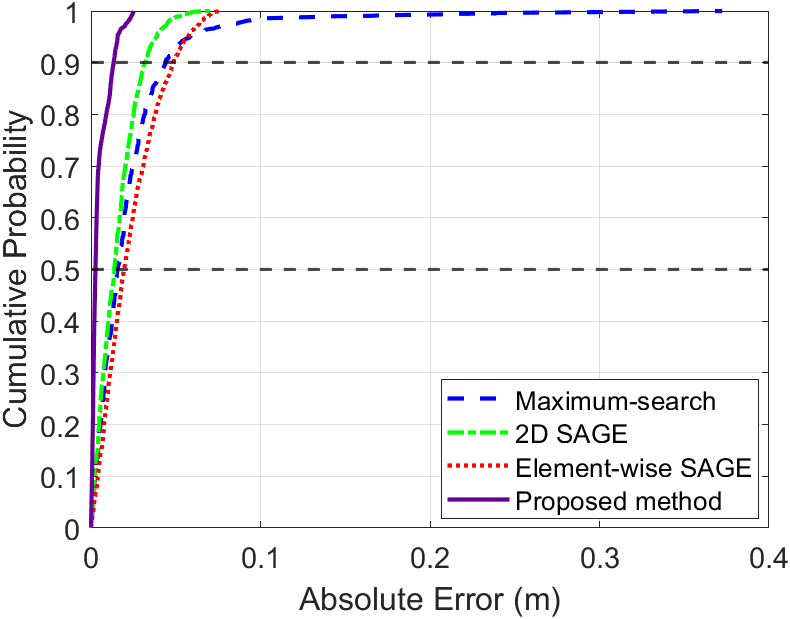}
\captionsetup{font={footnotesize}}
\caption{The CDF curve for ranging errors with respect to different methods.}
\label{fig:cdf_geometry}
\end{figure}

\begin{table}[t]
\captionsetup{font={footnotesize}}
\caption{Ranging error of different methods.}
\label{tab:error}
\centering
\begin{tabular}{{ccc}}
\toprule
Method                               & MDE (cm)  & RMSE (cm) \\
\midrule
Maximum-search algorithm             & 2.25 & 3.76 \\
2D SAGE in~\cite{fang2024centimeter} & 1.57 & 1.97 \\
Element-wise SAGE in~\cite{mpc_estimation} & 2.33 & 2.89 \\
Proposed algorithm                  & 0.49 & 0.73 \\
\bottomrule
\end{tabular}
\end{table}

Fig.~\ref{fig:recon_envir} presents the mapping results of a consistent wall structure, comparing three conventional algorithms. The maximum-search algorithm identifies the strongest path at each scanning direction and maps the corresponding reflection point. As shown in Fig.~\ref{fig:recon_envir}~(a), this method captures the general layout but fails to detect the corner due to the influence of the antenna radiation pattern. The 2D SAGE algorithm, as shown in Fig.~\ref{fig:recon_envir}~(b), refines the reconstruction by jointly estimating both delay and angle parameters. Fig.~\ref{fig:recon_envir}~(c) shows the performance of the element-wise SAGE algorithm, where the corner effect is mitigated, as it avoids overfitting to the distorted angular distributions caused by the antenna pattern. Nevertheless, noise and scattered paths remain visible in non-structural regions. In contrast, the proposed method, as shown in Fig.~\ref{fig:recon_new}, integrates CCA-based region segmentation and identification with a sliding-window refinement strategy, which allows for selective estimation of dominant MPC regions while preserving structural consistency. As a result, the reconstructed wall geometry exhibits high alignment with the ground truth. The ranging errors and cumulative distribution functions (CDFs) of different algorithms are presented in Table~\ref{tab:error} and Fig.~\ref{fig:cdf_geometry}, respectively. Thanks to the ultrawide bandwidth of 20~GHz, all methods achieve centimeter-level accuracy. The maximum-search algorithm, a simple and straightforward method that extracts the maximum power at each rotation angle, achieves a mean distance error (MDE) of 2.25~cm. However, it loses MPC information and fails to detect corners. The 2D SAGE algorithm proposed in~\cite{fang2024centimeter} estimates the parameters of each MPC and incorporates a geometric method to address corner effects, achieving an MDE of 1.57~cm. Nevertheless, mismatches in the antenna pattern lead to estimation inaccuracies within the processing window. The element-wise SAGE algorithm in~\cite{mpc_estimation} employs an MPC trajectory tracking strategy to mitigate antenna pattern effects across adjacent rotation angles, reaching an MDE of 2.33~cm. Building upon the SAGE approach in~\cite{mpc_estimation}, the proposed method integrates connected component region detection, structure identification, and a sliding-window strategy, substantially reducing the MDE to 0.49~cm and achieving millimeter-level reconstruction accuracy. Furthermore, the CDF results show that the proposed method achieves the lowest error distribution, with over 90\% of the points exhibiting errors below 1cm. These results confirm that the proposed CCA-assisted SAGE algorithm, combined with the sliding-window strategy, enables highly precise and robust environment reconstruction.

\begin{figure}[t]
\centering
\includegraphics[width=1.0\columnwidth]{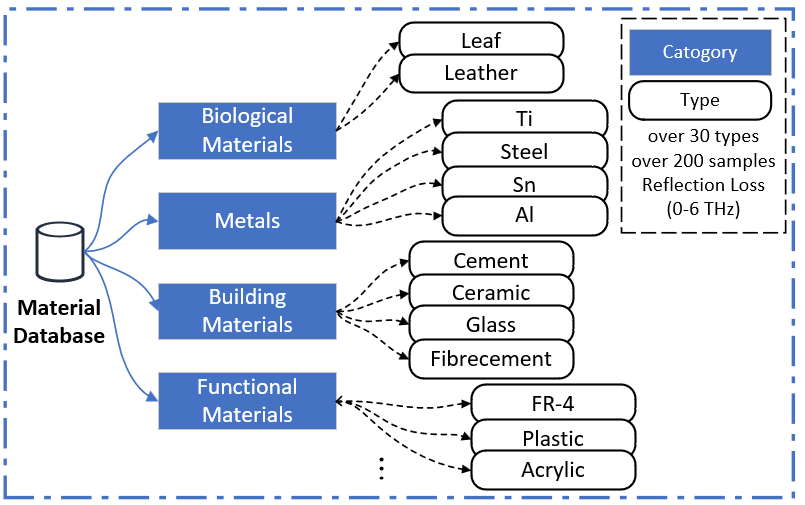}
\captionsetup{font={footnotesize}}
\caption{Material database measured by THz-TDS.}
\label{fig:database}
\end{figure}

\begin{table}[t]
\centering
\captionsetup{font={footnotesize}}
\caption{Exemplary reflection loss of different materials at 300~GHz.}
\label{tab:loss}
\begin{tabular}{cc}
\toprule
Material      & Reflection loss at 300~GHz (dB) \\
\midrule
Metal (e.g. Ti and Steel)            & 1.74-2.87                         \\
Cement        & 11.99                         \\
Ceramic       & 12.25                         \\
Fiber cement  & 13.24                         \\
Cardboard    & 17.00                         \\
Wood          & 20.56                         \\
\bottomrule
\end{tabular}
\end{table}

\begin{figure}
\centering
\subfigure[]{
\includegraphics[width=0.43\textwidth]{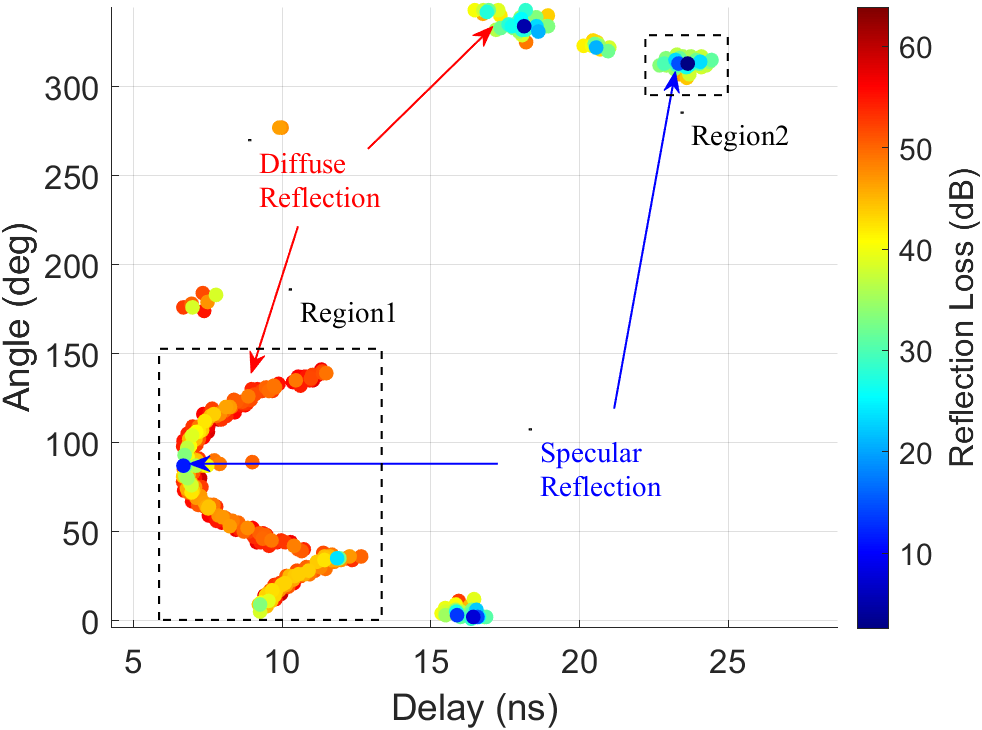}
\label{fig:RL_padp_wall}
}
\subfigure[]{
\includegraphics[width=0.43\textwidth]{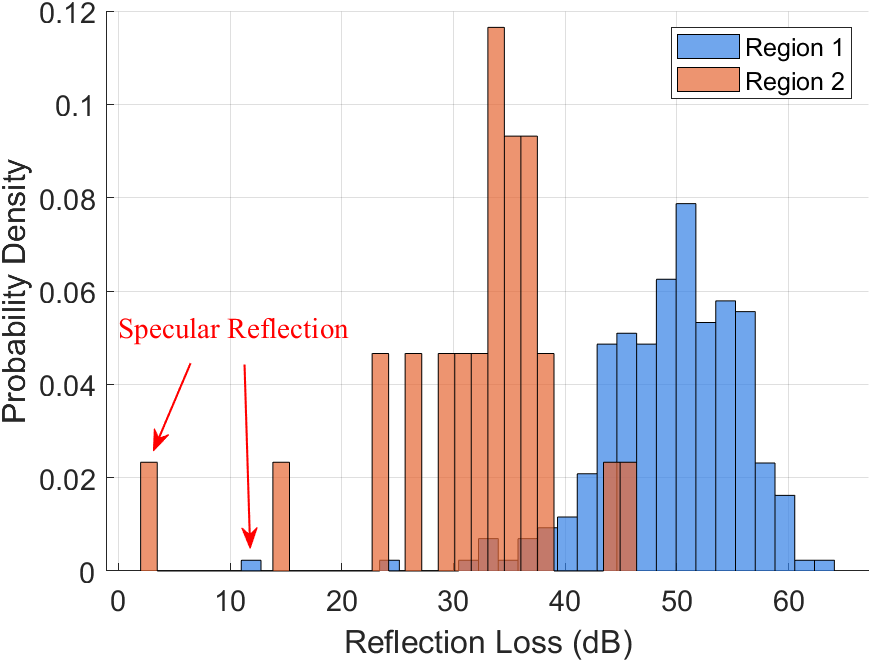}
\label{fig:RL_pdf_wall}
}
\captionsetup{font={footnotesize}}
\caption{Reflection loss analysis. (a) PADP of reflection loss at Region 1 and 2; (b) Distributions of reflection loss at Region 1 and 2. }
\label{fig:RL_wall}
\end{figure}

\subsection{Material Identification}
The material reflection database is established based on our own measurements through THz-TDS, as described in section~IV-B. As shown in Fig.~\ref{fig:database}, a database comprising over 30 types and over 200 common material samples is built, which is classified according to the physical and electromagnetic properties of different materials. Reflection losses of several typical indoor materials at 300~GHz are presented in Table~\ref{tab:loss}, including steel, cement, wood, etc. To validate the material identification approach, two representative regions are selected from the PADP segmentation, as highlighted in Fig.~\ref{fig:PADP_CV}. For each region, the reflection losses are computed for all estimated MPCs after SAGE. The calculated reflection loss values are then visualized as PADP heatmaps, as shown in Fig.~\ref{fig:RL_wall}~(a). Additionally, probability density functions (PDFs) of reflection loss are computed for both regions to quantify their reflection characteristics. As visualized in Fig.~\ref{fig:RL_wall}~(b), the reflection loss distribution for each region typically exhibits distinct separation between specular and diffuse reflections. The minimum reflection loss value within each region corresponds to specular reflection and directly used for comparison with the material database. For Region 1,  the PADP exhibits a combination of deterministic and stochastic reflections, indicating both specular and diffuse MPCs. This behavior is characteristic of rough-surface materials, where surface irregularities lead to scattering effects. The minimum reflection loss in this region is 11.40~dB, which matches closely to the reflection loss of cement, e.g., 11.99~dB, in the material database shown in Table~\ref{tab:loss}.
In contrast, for Region 2, the PADP displays highly localized reflections with minimal angular and delay dispersion, consistent with the characteristics of metallic surfaces. This region is indicative of strong specular reflection. The minimum reflection loss in this region is 2.50~dB, which corresponds closely to the value of metal, e.g., 1.74-2.87~dB.
The identification results are consistent with the materials of the wall and the metal structure of the window frame, respectively, validating the system's sensing ability to perform robust material identification based on a material database of reflection characteristics at 300~GHz.
\section{Conclusion}
In this paper, we present a computationally efficient framework for environment reconstruction and material identification in THz monostatic sensing. Experiments are conducted in an indoor scenario using a VNA-based channel sounder operating in the 290–310~GHz band. To address the computational issue in conventional SAGE estimation, an CCA-assisted SAGE-based MPC estimation algorithm is introduced, integrating closing operation and CCL algorithm to extract dominant multipath regions, leading to an 8.4$\times$ speedup in parameter estimation while maintaining high resolution. Moreover, the extracted connected component structures are analyzed to identify key indoor features such as flat walls and corners. A sliding-window filtering technique is then applied to suppress mapping distortions and enhance structural consistency. 20~GHz system bandwidth gives a 1.5~cm distance resolution, beyond which our CCA-assisted SAGE algorithm further refines to 4.9~mm, at least one order of magnitude better than that reported in the literature. In addition, the deterministic and stochastic components of the monostatic reflections are classified based on their spatial structure. The reflection losses of dominant surfaces are estimated at 300~GHz and matched with a material database comprising over 200 common samples based on THz-TDS. Two key structures are identified as cement and metal, respectively. These results highlight the potential of integrating the image processing technique with THz ISAC systems for enhanced environmental sensing. Future work will focus on extending the proposed framework by incorporating multi-modal information for material identification, thereby further improving the reliability and accuracy of environment reconstruction in real-world ISAC applications. 

The demonstration video of our database can be found at our lab website: \href{https://twclabsjtu.github.io/THz-TDS-Demo-Material-Database}{https://twclabsjtu.github.io/THz-TDS-Demo-Material-Database}.

\bibliographystyle{IEEEtran}
\bibliography{reference}

\end{document}